\documentclass[12pt,preprint]{aastex}

\usepackage{amsmath}
\usepackage{natbib}

\usepackage{graphicx}
\usepackage[figuresright]{rotating} 
\usepackage{float}       
\usepackage{afterpage}   

\newcommand {\IN}{$\Gamma$}
\newcommand {\SO}{H1743$-$322}

\newcommand{\wsim}{\ensuremath{\sim}}
\newcommand{\rxte}{\emph{RXTE}}
\newcommand{\tin}{$T_{in}$}
\newcommand{\integral}{\emph{INTEGRAL}}
\newcommand{\chandra}{\emph{Chandra X-ray Observatory}}

\begin{document}


\title{The Galactic black hole transient H1743$-$322 during outburst decay:
connections between timing noise, state transitions and radio emission}

\author{E. Kalemci\altaffilmark{1,2},
        J. A. Tomsick\altaffilmark{3},
        R. E. Rothschild\altaffilmark{3},
        K. Pottschmidt\altaffilmark{3},
        S. Corbel\altaffilmark{4},
        P. Kaaret\altaffilmark{5}
}

\altaffiltext{1}{Space Sciences Laboratory, 7 Gauss Way, University of
California, Berkeley, CA, 94720-7450, USA}

\altaffiltext{2}{Current address: Sabanc\i\ University,
Orhanl\i-Tuzla 34956, \.Istanbul, Turkey}

\altaffiltext{3}{Center for Astrophysics and Space Sciences, Code
0424, University of California at San Diego, La Jolla, CA,
92093-0424, USA}

\altaffiltext{4}{AIM - Unit\'e Mixte de Recherche CEA - CNRS -
Universit\'e Paris VII - UMR 7158, CEA Saclay, Service
d'Astrophysique, F-91191 Gif sur Yvette, France}

\altaffiltext{5}{Department of Physics and Astronomy, University of
Iowa, Van Allen Hall, Iowa City, IA, 52242, USA}


\begin{abstract}

Multi-wavelength observations of Galactic black hole transients during
outburst decay are instrumental for our understanding of the accretion
geometry and the formation of outflows around black hole systems. \SO, a black
hole transient observed intensely in X-rays and  also covered in the radio
band during its 2003 decay, provides clues about the changes in accretion
geometry during state transitions and also the general properties of X-ray
emission during the intermediate and the low-hard states. In this work, we
report on the evolution of spectral and temporal properties in X-rays and the
flux in the radio band with the goal of understanding the nature of state
transitions observed in this source. We concentrate on the transition from the
 thermal dominant state to the intermediate state that occurs on a timescale
of one day. We show that the state transition is associated with a
sudden increase in power-law flux. We determine that the ratio of
the power-law flux to the overall flux in the 3--25 keV band must
exceed 0.6 to observe strong timing noise. Even after the state
transition, once this ratio was below 0.6, the system transited back
to the thermal dominant state for a day. We show that the emission
from the compact radio core does not turn on during the transition
from the thermal dominant state to the intermediate state but does
turn on when the source reaches the low-hard state, as seen in
4U~1543$-$47 and GX~339$-$4. We find that the photon index
correlates strongly with the QPO frequency and anti-correlates with
the rms amplitude of variability. We also show that the variability
is more likely to be associated with the power-law emission than the
disk emission.

\end{abstract}

\keywords{black hole physics -- X-rays:stars -- accretion, accretion disks -- binaries:close --
stars:individual (H1743$-$322)}



\section{Introduction}\label{sec:intro}

The Galactic black hole transients  show several correlated spectral
and temporal variability properties during outbursts, denoted as
spectral states. During the initial rise and at the end of the decay
before quiescence, these transients are usually in the ``low-hard''
state (LHS). In this state, a hard power-law component dominates the
X-ray spectrum, and strong variability ($>$ 20\% rms amplitude) and
quasi-periodic oscillations (QPOs) are often observed. In between
the rise and the decay, the source may evolve through a combination
of ``thermal dominant'' and ``steep power-law'' states. In the
thermal dominant state (TDS), the soft disk component dominates the
spectrum, and the timing noise is very low or absent. In the steep
power law state, the power-law flux in the 2--20 keV band accounts
for more than 50\% of the flux and has a photon index (\IN) greater
than 2.4. Moderate variability and QPOs are observed in this state.
There also exist intermediate states (IS), where source
characteristics do not fit into the steep power law, TDS, or LHS,
but show various combinations of these states \citep[see][for
detailed discussion of spectral states]{McClintock03}. Throughout
this work, we will use the name IS for the particular intermediate
state between the TDS and the LHS during the outburst decay. Even
though these states were historically characterized using X-ray
observations, changes in other bands occur as well. In the TDS, the
radio emission from the compact core is quenched
\citep{Fender99,Corbel00}. Optically thin outflows are sometimes
detected during state transitions \citep{Fender01_c,Corbel01}, and
powerful, compact jets are always observed in the LHS
\citep{Fender01b}. The optical and infrared emission also show state
dependent properties \citep{Kalemci05,Homan05_a,Corbel02}.

The multi-wavelength observations made during the decaying portion of the
outbursts provide valuable information about black hole transients because of
the very high probability of observing transitions from the TDS to
 the IS, and eventually to the LHS \citep{Kalemci_tez}. The changes
during the transitions can reveal the geometry and the physical environment of
these systems before and after the transitions \citep{Esin97,Zdziarski02_2}.
The LHS contains additional information due to strong variability, and
strong radio emission, both correlating with spectral parameters. Our group
has been observing these transients during outburst decay in X-rays with the
 \emph{Rossi X-ray Timing Explorer} (\rxte) and in radio to understand the
 evolution before, during and after the state transitions
\citep{Kalemci01,Tomsick01b,Kalemci02,Tomsick03_2,Kalemci05}. Our emphasis
is on state transitions, and especially on understanding the changes in X-rays
while the radio jet is turning on. A uniform analysis of all black hole
transients observed with \wsim daily coverage with \rxte\ during outburst decay
 between 1996 and 2001 provided important information on the evolution of
spectral and temporal parameters during the decay \citep{Kalemci03}.
The sharpest change indicating a state transition is observed to be
a jump in the rms amplitude of variability from less than a few
percent to more than tens of percent in less than a day. This change
in the rms amplitude is almost always accompanied by a sharp
increase in the power-law flux. There is also evidence that the
strong rms noise is only observed when the power-law flux from the
source is above a certain percentage of the total flux. This sharp
change in rms amplitude of variability is noted as the time of state
transition from the TDS to a harder state in \cite{Kalemci03}, and
the same definition will be applied here. During the outburst decay,
the photon index, the disk temperature, and the disk flux usually
decrease slowly. Often, late in the outburst, the disk flux becomes
undetectable. After the transition, characteristic frequencies of
the power density spectrum also decrease with time.

 \SO\ was discovered with \emph{Ariel 5} \citep{Kaluzienski77_iauc1}
and \emph{HEAO-1} \citep{Doxsey77_iauc} satellites in August 1977.
After a couple of detections in 1984 with \emph{EXOSAT}
\citep{Reynolds99}, and in 1996 with TTM/COMIS on \emph{Mir-Kvant}
\citep{Emelyanov00}, the source was detected in outburst again in
March 2003 with \integral\ \citep{Revnivtsev03_atel}, and \rxte\
\citep{Markwardt03_atel}. The radio \citep{Rupen03_atel}, infrared
\citep{Baba03_iauc} and optical \citep{Steeghs03_atel} counterparts
were quickly identified during the 2003 outburst. The radio
observations revealed relativistic jets \citep[$v/c \eqsim
0.8$,][]{Rupen04, Corbel05}. Large scale jets were also detected in
X-rays with the \chandra\ at the end of the outburst
\citep{Corbel05}. The X-ray observations with \rxte\ and \integral\
indicate that the source went through several spectral states before
fading at the end of 2003 \citep{Markwardt03_atel, Homan03_atel,
Kretschmar03_atel, Grebenev03_atel, Tomsick03_atel, Parmar03,
Joinet05}. Even though there is no mass measurement of the compact
object, the X-ray spectral and temporal properties, and a high
frequency QPO pair with frequencies similar to those of other black
hole sources \citep{Homan05} establish this source as a very likely
black hole.


In this work, we will characterize the X-ray and radio properties of \SO\
during the outburst decay in 2003, compare these properties to the general
properties of black hole transients, and discuss the unique properties of this
source in detail. We will especially concentrate on the triggering mechanism
for the state transitions during the outburst decay.


\section{Observations and Analysis}\label{sec:obs}

\subsection{The RXTE observations}\label{subsec:rxteobs}

\begin{figure}
\plotone{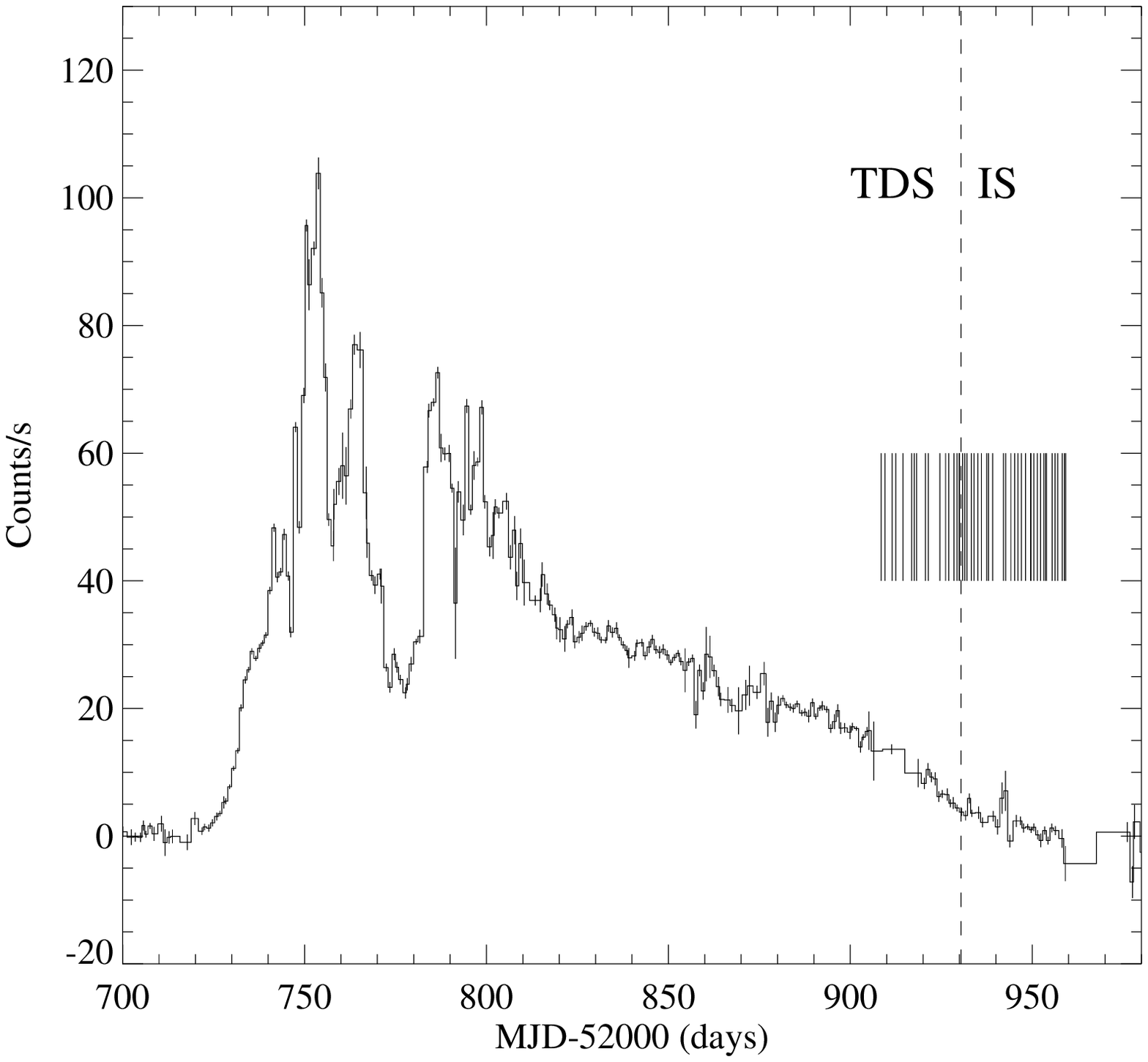}
\caption{\label{fig:asm}
The 2--12 keV ASM light curve of the 2003 outburst of \SO. The vertical lines
indicate the times for \rxte\ pointings that we analyzed. The dashed line
indicates the approximate time of state transitions from the TDS to the
IS.
}
\end{figure}

The \rxte\ ASM light curve of \SO\ in the 2003 outburst is shown in
Fig.~\ref{fig:asm}. Our daily monitoring campaign with \rxte\ (under
observation ID 80137, each pointing 2-3 ks long) started on MJD~52906 after
the source's ASM count rate dropped below 15 cts/s. The source was in the TDS
until MJD~52930 when a transition to the IS occurred \citep{Tomsick03_atel}.
After MJD~52938, the source was in the LHS \citep{Kalemci04_T}. In the LHS, we
also obtained longer exposure observations (10-15 ks, see
Table~\ref{table:spe_par}) to investigate spectral and temporal properties of
the source in greater detail than the daily monitoring observations. The
monitoring program has very good coverage of the transitions and evolution
during the outburst decay (see Fig.~\ref{fig:asm}). Table~\ref{table:spe_par}
shows the list of the observations we used in this work.


\subsubsection{X-ray spectral analysis}\label{subsubsec:rxtespec}

For most of the observations, we used both the PCA and the HEXTE
instruments on \rxte\ for the spectral analysis (see \citealt{Bradt93} for
instrument descriptions). For the PCA, the 3--25 keV band
was used, and the response matrix and the background model were created
using the standard FTOOLS (version 5.3.1) programs. We added 0.8\% up to 7 keV,
 and 0.4\% above 7 keV as systematic error \citep[for the details of
how we estimated systematic uncertainties, see][]{Tomsick01b}. We used all
available PCUs for each observation, choosing the combination that would give
maximum number of counts per pointing.

The 15--200 keV band was used for the HEXTE data. We used the response matrix
created by the FTOOLS, and applied the necessary deadtime correction
\citep{Rothschild98}. The HEXTE background is measured throughout the
observation by alternating between the source and background fields every
32 s. As \SO\ is close to the plane ($b$=-1.83$^{\circ}$), both the source
confusion and the Galactic ridge emission are important for determining the
HEXTE background. Using the \emph{HEXTEROCK} utility and Galactic bulge scans,
and by comparing the rates between ``+'' and ``-'' off-source cluster
positions, we determined that for both clusters, ``+'' background pointing
provides a better estimate of the background for our observations. For Cluster
A, the ``-'' pointing is in the Galactic plane with very strong ridge
contribution, and background sources are contributing in Cluster B ``-''
position. The relative normalization between the PCA and the HEXTE is kept
free, and varies between 0.9 and 1. The HEXTE data was included in the
spectral analysis until MJD~52950. After this date, including HEXTE data did
not improve the fits due to low fluxes and short observation times.

The Galactic ridge emission was a factor for all of the observations discussed
here because of the proximity of the source to the Galactic plane. We
determined that our last 9 observations in 2004 (MJD~53021--53055) had a
constant flux that represented the ridge emission with a power-law index of
2.325 and a Gaussian at 6.62 keV. The 3--25 keV unabsorbed flux from the ridge
emission is 1.08 $\times$ 10$^{-10}$ ergs cm$^{-2}$ s$^{-1}$. These numbers
are consistent with expectations from Galactic ridge \citep{Revnivtsev03}. To
make sure, we also analyzed the \rxte\ Galactic bulge scans before and after
the outburst. We determined that the model described above fits them as well.
For all observations, we fixed the ridge parameters as given above. After
MJD~52960, our monitoring observations did not detect the source significantly
above the Galactic ridge emission.

For all the observations, our first spectral model consisted of
absorption (``phabs'' in XSPEC), smeared edge \citep[``smedge'' in
XSPEC,][]{Ebisawa94}, a multicolor disk blackbody \citep[``diskbb''
in XSPEC,][]{Makishima86}, a power law (``pegpwrlw'' in XSPEC), and
a narrow Gaussian to model the iron line, and the ridge emission (as
described above). This model has been commonly used for the spectral
analysis of black holes in the LHS
\citep{Tomsick00,Sobczak00,Kalemci05}. The hydrogen column density
was fixed to $\rm N_{H} = 2.3 \times 10^{22} \, cm^{-2}$, based on
the \chandra\ results \citep{Miller04}. The smeared edge width was
fixed to 10 keV. Once we fit the observations with this model, we
added a high energy cut-off (``highecut'' in XSPEC) to the model.
None of the observations showed a significant decrease in $\chi^{2}$
that would indicate a high energy cut-off.

\subsubsection{X-ray temporal analysis}\label{subsubsec:rxtetim}

For each observation, we computed the power density spectra (PDS)
from the PCA data using IDL programs developed at the University of
T\"{u}bingen \citep{Pottschmidt02th} for three energy bands, 3 -- 6 keV, 6 --
15 keV, and 15 -- 30 keV. We also computed the PDS for the combined
band of 3 -- 30 keV. The source flux above 30 keV is too low for timing
analysis. The PDS was normalized as described in \cite{Miyamoto89} and
corrected for the dead-time effects according to \cite{Zhang95} with a
dead-time of $\rm 10\,\mu s$ per event. Using 256 second time segments, we
investigated the low frequency QPOs and the timing properties of
the continuum up to 256 Hz for different energy bands. We fit all our PDSs
with broad and narrow Lorentzians with our standard timing analysis techniques
 \citep{Kalemci05,Kalemci_tez,Pottschmidt02th}.

In the PDS fits, Lorentzians with quality value (centroid frequency divided by
the width) $Q$ $>$ 2 are denoted as QPOs. The
rms amplitudes are calculated over a frequency band from zero to infinity. We
multiplied the rms amplitude of variability with ${T^{2}}\over{(T-(R+B))^{2}}$,
where $T$ is the overall count rate, $B$ is the background rate determined
using \emph{pcabackest}, and $R$ is the count rate due to the Galactic ridge,
to obtain the variability inherent to the source \citep{Berger94}.

\subsection{Radio observations}\label{subsec:radio}

We obtained \emph{Very Large Array} (VLA) radio information (observations
conducted by M. Rupen) for dates between MJD~52920 and MJD~52949. The
``core'' was not detected in 4.86 and 8.64 GHz frequencies on MJD~52924.9  and
MJD~52933.0. Another radio component (from an earlier discrete plasma
ejection) away from the core was detected during this time
\citep{Rupen04_atel2}. The first detection of the core during the decay
occurred on MJD~52940.0, with a flux density of 0.14$\pm$0.04 mJy/beam (M.
Rupen, personal communication). The last detection of the core with the VLA
occurred on MJD~52949, with a flux density of 0.22$\pm$0.04 mJy at 4.86 GHz
\citep{Rupen04_atel1}.

\begin{figure}
\plotone{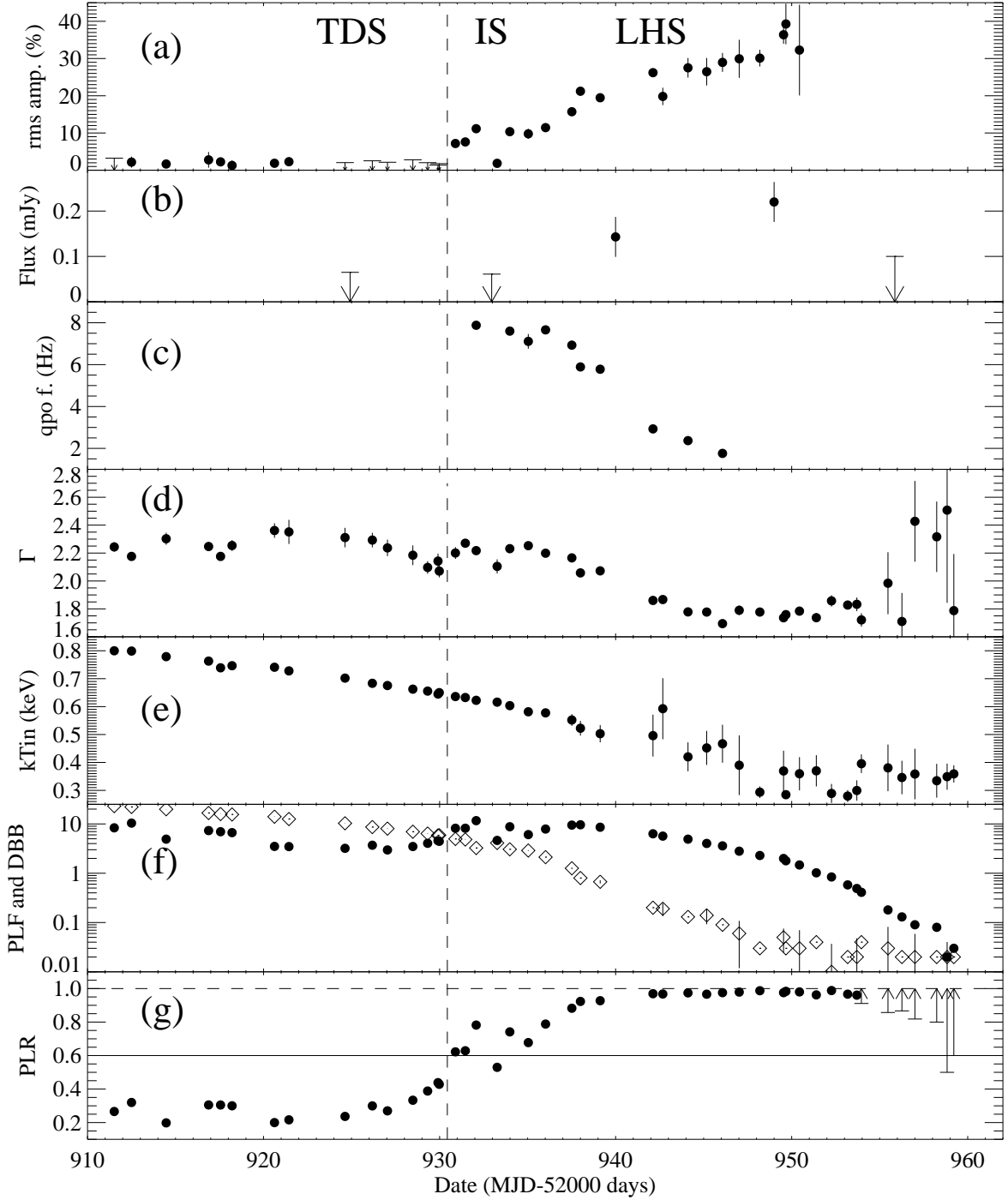}
\caption{\label{fig:evol1}
The evolution of (a) the total rms amplitude of variability in 3-30 keV band,
(b) the radio flux density at 4.86 GHz (c) the QPO frequency, (d) the photon
index (\IN), (e) the inner disk temperature \tin\, (f) the power-law flux
(circles) and the disk-blackbody flux (diamonds) in 3--25 keV band in units of
10$^{-10}$ ergs cm$^{-2}$ s$^{-1}$, (g) the ratio of the power law flux to the
total flux in 3--25 keV band. The dashed line indicates the case for which all
the emission comes from the power-law component. The solid line shows the
threshold for observing timing noise. Unabsorbed fluxes are shown.
For most of the measurements, the 1 $\sigma$ uncertainties are smaller than
plot symbols. The vertical dashed line indicate the approximate time of
transition from the TDS to the IS. The first four points in panel (b) are
from VLA, and the final point is from ATCA.
}
\end{figure}

Towards the end of the 2003 outburst, the source was observed with
ATCA. The source was not detected on MJD~52955.86, 52973.72, 52983.45,
52994.49, 53049.40, with the highest rms value of 0.10 mJy at 4.8 GHz
\citep{Corbel05}. The evolution of the radio fluxes
during the outburst decay is given in Fig.~\ref{fig:evol1}.b.



\section{Results}\label{sec:results}

We investigated the evolution of several spectral and temporal fit
parameters during the decay of \SO\ to establish the time of state
transitions. Some of the important parameters are shown in
Fig.~\ref{fig:evol1} and some of the spectral fit parameters are also
tabulated in Table~\ref{table:spe_par}.


\subsection{States and transitions}\label{subsec:states}

\begin{figure}[t]
\plotone{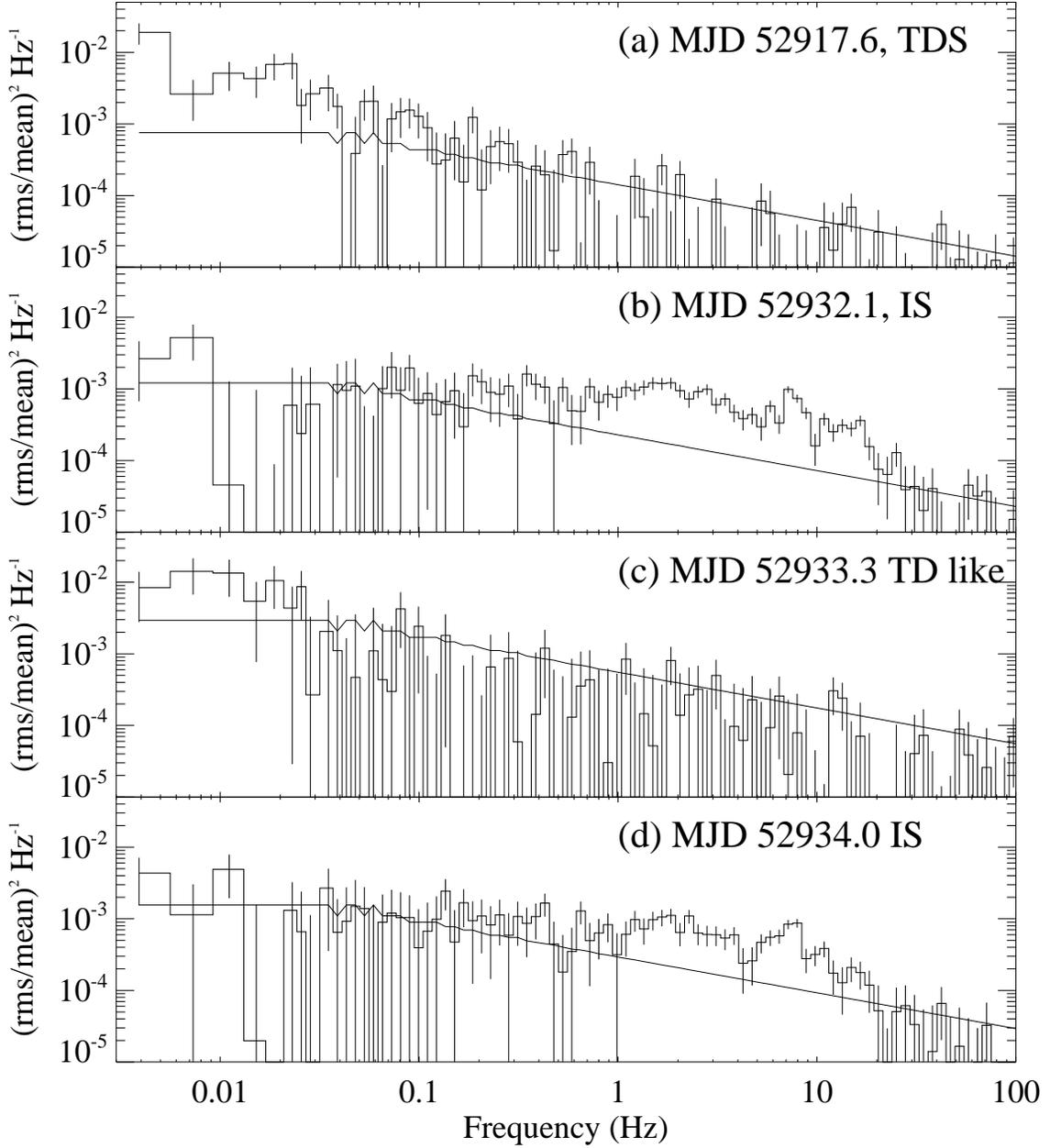}
\caption{\label{fig:psds}
Power spectra in (a) the TDS, (b) after the transition to the IS, (c)
during the TD-like observation, (d) right after the TD-like observation, again
in the IS. PDSs in (b), (c), and (d) are approximately a day apart. The solid
line shows the effective Poisson noise level \citep[uncertainty in Poisson
noise subtraction after merging segments and binning the
data,][]{Nowak99}.
}
\end{figure}

Until MJD~52930, the spectral and temporal properties of the source indicate
a TDS. When the noise was detected in the PDS, it was characterized by
a weak power-law typically observed in the TDS
(see Fig.~\ref{fig:psds}.a). The rms amplitude of variability was less than
3\%.  On MJD~52930, the power-law flux doubled while the  power-law ratio
(PLR, the ratio of the power-law flux to the total flux in 3--25 keV band)
exceeded 0.6, and the PDS showed detectable timing noise with an rms
amplitude of \wsim 7\%. The spectral and temporal properties (see
Fig.~\ref{fig:psds}, b and d) indicate an IS right after this transition.
During the transition, the photon index increased slightly. The evolution of
the $\rm T_{in}$ and the disk-blackbody (DBB) flux were smooth.

A QPO appeared on MJD~52932 as the PLR reached 0.8.  Just after this
date, for one observation, the power-law flux and the PLR dropped.
The strong timing signature also disappeared, and the PDS only
showed a weak power-law component. The source properties were
similar to those of the TDS. This single observation will be called
``TD-like'' for this reason. Within half a day, the power-law
 flux increased to the IS levels, and the timing noise, along with the QPO
reappeared (see Fig.~\ref{fig:psds}). Based on the TDS observations
 and the TD-like observation, we set a PLR threshold of variability of
\wsim 0.6 for \SO. The duration for the TD-like case is of the order
of a day. Interestingly, the photon index decreased during the
TD-like observation, returning to the value measured right before
the transition to the IS.

QPOs with decreasing frequency were observed in ten observations after the
transition. After MJD~52933, the rms amplitude of variability increased, the
photon index hardened, and the inner disk temperature and the DBB flux
decreased smoothly. At the end of this evolution, the source was in the LHS.
 The transition to the LHS was smooth, and it is hard to determine
when the source left the IS. Based on the \cite{McClintock03}
criteria (spectrum dominated by the power-law component, 1.5 $<$
\IN\ $<$2.1, rms amplitude of variability $>$ 10\%), the transition
happened around MJD~52938. In the LHS the photon index hovered
between 1.7 and 1.9 until MJD~52956. The rms amplitude of
variability increased until MJD~52950. After this date, the
statistical quality of the data was not good enough to determine the
rms amplitude of variability.

\subsection{Correlations between spectral and temporal parameters}

\begin{figure}
\plotone{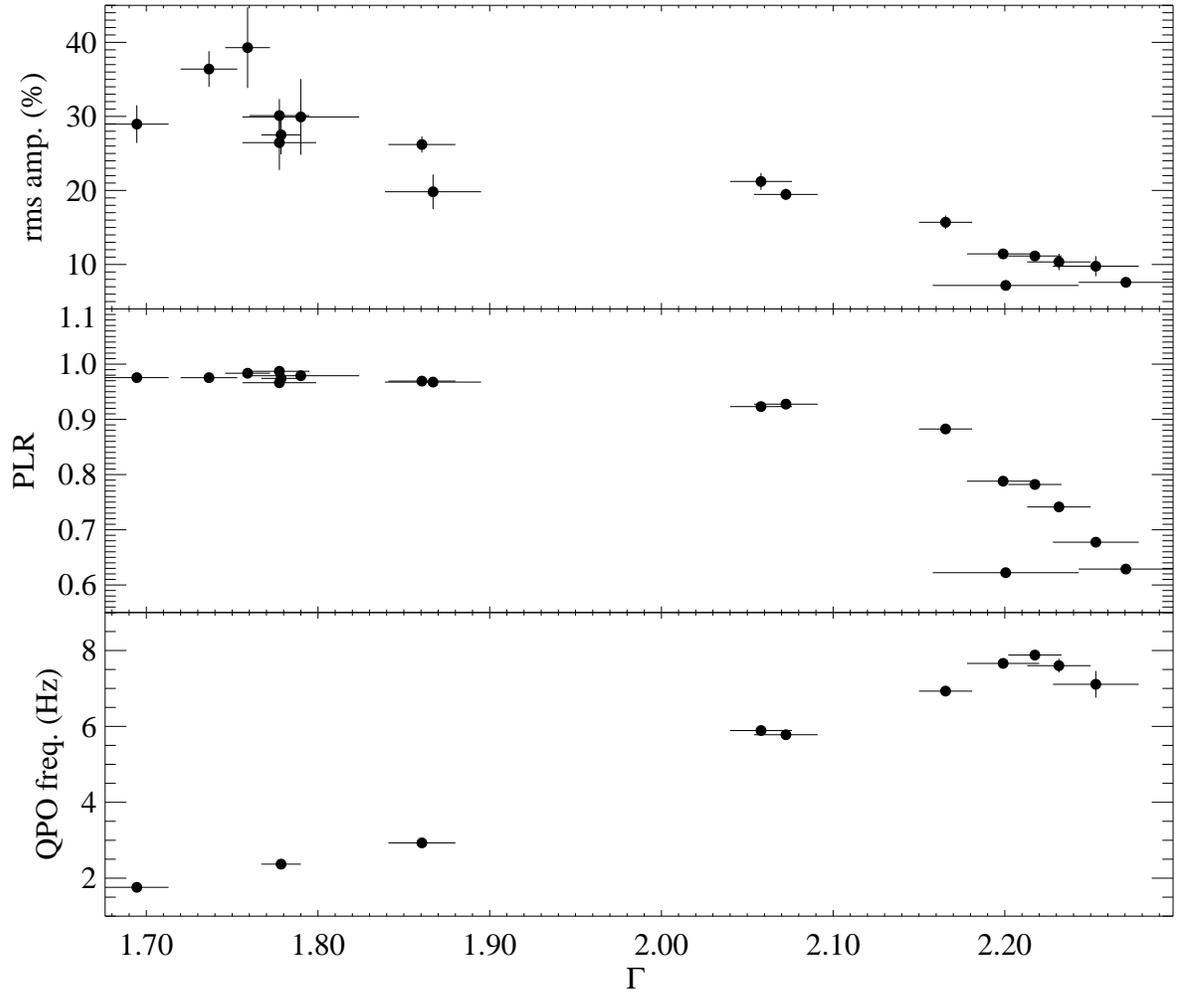}
\caption{\label{fig:indcor}
The correlation between the photon index and, (a) the rms amplitude of
variability, (b) the PLR, and (c) the QPO frequency for observations
taken between MJD~52930.9 and MJD~52949.7, excluding the TD-like observation.
}
\end{figure}

It has been well established that the black hole transients show
several correlations between spectral and temporal parameters
\citep[][and references therein]{Kalemci_tez}. During outburst
decays, the photon index shows the strongest correlations (or
anti-correlations) with the QPO frequency and the rms amplitude of
variability \citep{Vignarca03,
 Kalemci_tez}. We plotted these parameters as a function of
photon index in Fig.~\ref{fig:indcor}. We also included the PLR as a
function of photon index, as it shows a strong correlation with the
rms amplitude of variability in the IS. Fig.~\ref{fig:indcor} shows
that the rms amplitude of variability is strongly anti-correlated
(linear correlation coefficient of {-0.949}), and the QPO frequency
is strongly correlated (linear correlation coefficient of 0.987)
with the photon index. Notice that there might be a turnover (or a
saturation) in the QPO-\IN\ relation, as seen earlier for other
sources by \cite{Vignarca03}. The TD-like observation is not shown.
Obviously, it does not obey the anti-correlation between the rms
amplitude of variability and \IN.

\begin{figure}
\plotone{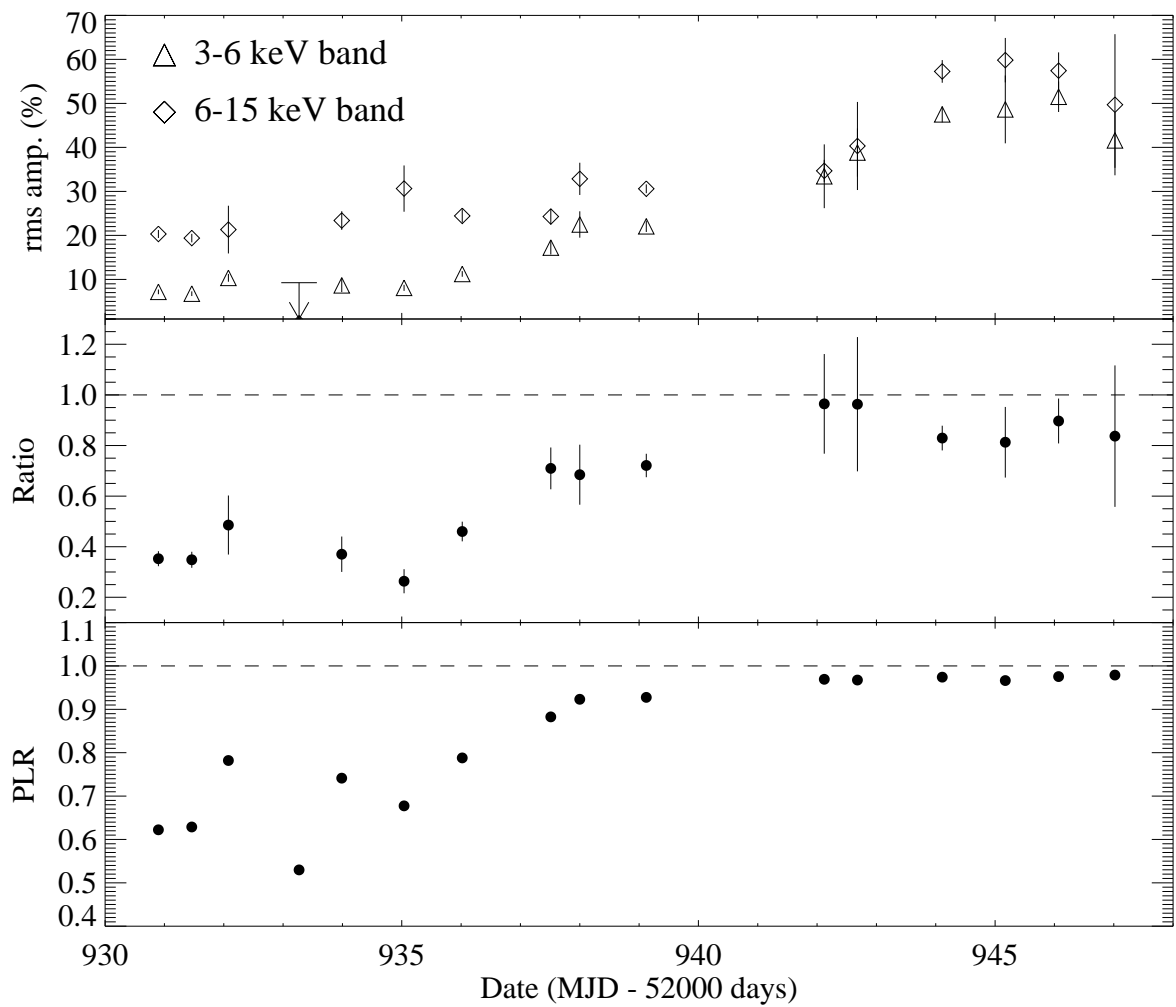}
\caption{\label{fig:rmsen}
The evolution of (a) the rms amplitude of variability in 3--6 keV (triangles)
and 6--15 keV band (diamonds). The 3$\sigma$ upper limit rms amplitude of
variability for the TD-like observation in 6--15 keV band (9.25 \%) is shown
with an arrow. The 3--6 keV band upper limit of 3.28 \% is not shown for
clarity. (b) The ratio of the rms variability in 3--6 keV
and 6--15 keV band, (c) the PLR.
}
\end{figure}

Another interesting relation is between the rms amplitude of
variability and the disk flux. It is well known that the rms
amplitude of variability increases with energy for systems with
significant disk emission \citep{Kalemci_tez}, but once this disk
component is not significant, the relation between rms amplitude of
variability and energy is not predictable \citep{Kalemci03}. In
Fig.~\ref{fig:rmsen}, we show the rms amplitude of variability for
two energy bands, their ratio and the PLR. The presence of DBB
emission clearly reduces the rms amplitude of variability in 3--6
keV band. For most of the observations during which the DBB emission
is not significant, the ratio between rms amplitudes of variability
is consistent with being unity. But when the last six observations
are grouped, it is clear that the 6--15 keV band rms amplitude of
variability is higher than that of the 3--6 keV band.



\section{Discussion}\label{sec:discussion}

\subsection{Evolution of spectral and temporal properties during outburst decay}

In general, the evolution of spectral and temporal parameters of
\SO\ show very similar characteristics to the evolution of other
black hole systems during outburst decay. The transition from the TDS to
the IS is sharp, and is marked by a jump in the rms amplitude of
variability, the power-law flux and the PLR \citep{Kalemci03, Kalemci05}. We
note that, during the transition, both the power-law flux and the PLR
increase substantially. But the sudden increase in the PLR is always due
to a sudden increase in the power-law flux, and not due to a sudden decrease in
 DBB emission. Therefore, the physical change that causes the state transition
is associated with a sudden increase in the power-law flux. On the other hand,
the PLR provides a threshold for strong variability. After the transition from
 the TDS to the IS, the DBB flux and the inner disk temperature steadily
decreases, and finally the source reaches the LHS.

A detailed discussion of several scenarios which could result in
such evolution during the decay is given for 4U~1543$-$47 in
\cite{Kalemci05}. \SO\ shows quite similar spectral and temporal
evolution in X-rays to 4U~1543$-$47. In addition, for both cases,
the radio core was only detected after the sources reached their
hardest levels, and not detected right after the transition to the
IS. For \SO, the fits never require a high energy cut-off in the
spectrum, while for 4U~1543$-$47, a cut-off is observed only during
the transition to the LHS. It is possible that a portion of
electrons are non-thermal for both sources, at least in the TDS and
IS \citep{Coppi00}. For \SO\, the evolution of \tin\ is consistent
with an accretion disk receding away from the black hole (and the
drop in QPO frequency may also be an indication of this).  The
evolution of the PLR and \IN\ indicate a Comptonizing medium (either
an independent corona or base of a jet) getting stronger as the
decay progresses. The lack of adequate multi-wavelength coverage
does not allow us to strongly constrain the details of different
models as we were able to do for 4U~1543$-$47.

Even though most of the properties of \SO\ are generic to the black
hole transients during decay, it also showed two unexpected
behaviors. The first one is the TD-like observation on MJD~52932.
Once the transition to a harder state occurs, it is unusual for a
system to go back to the previous, softer state just for a day
during the outburst decay. (Short branchings to harder states are
common, such as branching seen in the 1998 outburst of
XTE~J1550$-$564, \citealt{Homan01}). For the 14 outburst decays
analyzed in \cite{Kalemci_tez}, only XTE~J1859$+$226 showed a
similar behavior in its outburst decay in 1999. After the transition
to the IS at MJD~51524, XTE~J1859$+$226 transited back to the TDS
twice for a short time \citep{Kalemci_tez}. A recent report
indicates that during the decay of the 2002/2003 outburst of
GX~339$-$4, the source transited back to the TDS for three days
after the transition to the IS \citep{Belloni05}.  We note that
these branchings with time scales of a few days may be more common,
as frequent monitoring is required to catch them. If we examine the
cases source by source, rather than outburst by outburst, from the
seven transients that have been observed with daily monitoring,
three sources showed this behavior. The IS is a transitional state
between the TDS and the LHS, but these branchings imply that the
evolution is not monotonic. Transitions back and forth to the TDS
can occur on a time scale of days. In contrast, once the source is
in the hard state and the X-ray luminosity keeps decreasing, the
process is irreversible, there is no example (as best of our
knowledge) of a transition back to the IS or the TDS.

Second unexpected behavior is the slight increase in photon index  (\wsim 6\%)
during the transition to the IS (see Fig.~\ref{fig:evol1}). The only other
case that showed an increase in photon index during a transition from the TDS
 to the IS is the 2000 outburst of XTE~J1550$-$564
\citep{Kalemci03,Rodriguez03}. During such transitions, the power-law index
usually does not change, or decreases somewhat \citep{Kalemci03}. For both
\SO, and XTE~J1550$-$564, the power-law flux and the PLR showed a sudden
increase, while the photon index also increased. However, for \SO,  this
transition resulted in an increase in rms amplitude of variability, whereas
for XTE~J1550$-$564, it resulted in a decrease of rms amplitude of variability.
 Note that the conditions for these two transitions were not the same. For \SO,
 the transition was from TDS to an IS, whereas for XTE~J1550$-$564, the
transition was from a soft IS to a harder IS, en route to the LHS.

\subsection{Emission mechanisms during the state transition from the TDS to the IS}\label{sec:disst}

In the TDS, the bulk of the emission comes from the disk as
multi-color blackbody radiation. The evolution of the DBB component
during the transition to the IS is smooth. On the other hand, the
power-law component in the X-ray spectrum shows a sudden change, and
is perhaps associated with the state transition. For the power-law
component, the two important observations are (1) a sudden (less
than a day) and a strong (100\%) increase in the flux, and (2) a
small change  (6\% increase) in the photon index.  We note that an
increase in the power-law flux with only little or no change in the
power-law index is a universal property of the state transitions
that coincide with appearance of timing noise and QPOs for most of
the black hole systems \citep{Kalemci03}. In this section, we
discuss various scenarios that may explain the evolution of the
power-law component during the transition to the IS.

 For thermal Comptonization, the photon index depends on two parameters of the
corona: The temperature and the optical depth of the Comptonizing
medium \citep{Sunyaev80}. The power-law flux depends on the input
soft flux, the optical depth and the covering fraction of the
Comptonizing medium. Increasing the soft input flux would result in
a larger power-law flux and also may cause a lower coronal
temperature due to cooling. Our observations do not indicate an
increase in the DBB flux. A sudden expansion of the corona (while
keeping the optical depth constant) could explain the change in
power-law flux if the volume increase corresponds to an increase in
the covering fraction. Before the transition, the rms amplitude of
variability is less than 1.8\%, therefore there is at least a factor
of four increase in the rms amplitude of variability, whereas, there
is only a factor of 2 increase in the power-law flux. Therefore the
volume increase should have a non-linear relation to the variability
amplitude to explain the strong increase. Emergence of another soft
photon source inside the corona (with a spectrum not contributing in
the PCA range) is another possible explanation for the transition.
One candidate is synchrotron radiation inside the corona causing
synchrotron self-Compton \citep{Markoff04}. In this case, the hard
power-law emission will be created by a combination of thermal
Comptonization of synchrotron and disk photons and non-thermal
synchrotron self-Compton. For the case of synchrotron, the seed
photons come from inside, while for the DBB, the seed photons come
from both the outside and the inside of the corona, depending on the
assumed geometry. It is possible that the strong variability is
created more efficiently if the source is inside the corona. Since
the major change is not related to the temperature and the optical
depth of the corona, it could also explain why the photon index
varies little during the transition. The small change in the index
may be due to additional cooling (if the increase in the secondary
soft flux is more than the decrease in the DBB flux), or the effect
of additional non-thermal Comptonization. If the strong variability
in the IS and the LHS is due to an additional soft component such as
the synchrotron radiation, then the origin of low frequency QPOs
will not be the accretion disk \citep{Lee98,Titarchuk04}. One can
also envision a combination of scenarios, such as an expansion of
corona while the synchrotron self-Compton contribution is
increasing.

Another way to increase the power-law flux is to have two emission
mechanisms producing power-law spectra. During the transition, the
second component, which also produces the timing noise, may become
active. One possibility is that during the TDS the bulk motion
(dynamical) Comptonization \citep{Laurent99,Titarchuk04} may be the
main mechanism, whereas in the IS, thermal Comptonization starts to
operate. This requires that both components produce very similar
power-law spectra right at the transition. The secondary component
in the IS may also be direct synchrotron emission from a jet
\citep{Falcke99,Markoff01}. This, again, requires that both emission
mechanisms producing similar spectrum at the transition. There is
another argument against this second possibility. The radio
observations indicate that the emission from the compact jet began
sometime between MJD~52933 and MJD~52940, well after the transition
to the IS. (Since there were no optical and infrared monitoring of
the source during the transition, we were not able to pinpoint the
time of jet formation.) Based on the comparison between the optical
and IR light curves, X-ray and radio properties of GX~339-4
\citep{Homan05_a} and 4U~1543$-$47 \citep{Kalemci05}, one can argue
that the compact jet does not turn on before the source is settled
in the LHS. \SO\ seems to support this argument, as the compact
radio jet was first detected after the source was in the LHS
according to \cite{McClintock03}. We note that, during the rise, a
compact radio core was observed, even after the transition from the
LHS to the IS \citep{Joinet05,Rupen03_atel2}. This shows that, even
though the \emph{formation} of the compact jet requires the LHS with
almost no soft emission in the PCA band, it can be \emph{sustained}
for a while in the presence of soft emission \citep{Corbel04}.

\subsection{The correlations}\label{subsec:disccor}

There is a very strong correlation between the photon index and the QPO
frequency up to the photon index of 2.2. At high photon indices, the
relation seems to be turning over. XTE~J1550$-$564 and GRO~J1655$-$40 are two
other sources that show a turnover in QPO frequency - photon index relation
\citep{Vignarca03}. Relativistic  effects may cause this turnover, if the QPO
has an ``accretion ejection instability'' origin \citep{Varniere02,Vignarca03}.
 We note that we cannot conclusively claim that the relation turns over
for \SO, because the evidence for a turnover is based on a single data point
(see Fig.~\ref{fig:indcor}).  A saturation QPO frequency with photon index
could be a sign of a system dominated by a converging inflow
\citep{Titarchuk04}.

There is a strong anti-correlation between the photon index and the rms
amplitude of variability. This is a global relation that is valid for most
black hole transients during outburst decay \citep{Kalemci02}. One can also
observe a strong correlation between the PLR and the rms amplitude of
variability at high spectral indices (see Fig.~\ref{fig:indcor}). This shows
the diminishing effect of the DBB radiation on the rms amplitude of
variability. Another way to investigate the effect of DBB on the rms amplitude
 of variability is to compare the energy dependence of variability.
Fig.~\ref{fig:rmsen} shows that for the observations with strong DBB emission,
the rms amplitude of variability is much stronger in the higher energy band
than that of the lower energy band. One can see a correlation between the
ratio of the rms amplitude of variability in 2 bands and the PLR. Once the DBB
emission is negligible (PLR \wsim 1), the relation between the rms amplitude
 of variability and energy is not that clear. The correlations between the
PLR and both the rms amplitude of variability and the variability ratio
 indicate that the variability is generated in the power-law component, and
the DBB component regulates the rms amplitude of variability by changing the
mean flux \citep{Churazov01,Kalemci_tez}.

\section{Summary and Conclusions}

We analyzed the \rxte\ X-ray observations of \SO\ during its
outburst decay in 2003. In addition to the X-ray observations, we
also obtained radio fluxes and discussed their evolution with
respect to the state transitions. The evolution of the spectral and
temporal fit parameters show similar properties to those of other
sources. The transition from the TDS to the IS is marked by a strong
increase in the rms amplitude of variability and power-law flux.
This transition may correspond to the emergence of a secondary soft
component, such as synchrotron radiation, or a secondary power-law
component. Future multi-wavelength observations of these transients
may help making the distinction. At the time of this transition the
radio core was not detected. Three days after the transition, the
source went back to the TDS for one day, indicating that the IS is
not stable. After the source returned to the IS, it gradually
reached the low hard state. The core is detected in the radio band
when the X-ray spectrum is totally dominated by the power-law
emission.

  There is a strong correlation between the photon index and the QPO
 frequency. At high QPO frequencies and photon indices, the correlation
shows a turnover (or saturates). The threshold PLR for the state transition,
and the correlation between the PLR and the ratio of rms amplitude of
variability in  6--15 keV and 3--6 keV bands show that the DBB emission
 dilutes the amplitude of variability. When the DBB emission is
 absent, the rms amplitude of variability in two bands are similar.
 This may indicate that the origin of the variability and the QPO is the
 corona itself, rather than the disk.


\acknowledgments E.K. acknowledges NASA grant NAG5-13142 and partial
support of  T\"UB\.ITAK. E.K. thanks all scientists who contributed
to the T\"{u}bingen Timing Tools. The authors thank Michael Rupen
for the valuable VLA data. E.K. thanks Lev Titarchuk for his remarks
on state transition and evolution. This work made use of the
Galactic Bulge Scans of Craig Markwardt. J.A.T. acknowledges partial
support from NASA Grant NNG04GB19G. R.E.R. acknowledges NASA grant
NAS5-30720. P.K. acknowledges partial support from a University of
Iowa Faculty Scholar Award. The authors also thank the anonymous
reviewer for his comments that significantly improved the scientific
quality of the paper. The Australia Telescope is funded by the
Commonwealth of Australia for operation as a national Facility
managed by CSIRO.



\clearpage


\clearpage

\clearpage

\clearpage

\clearpage

\clearpage


\begin{sidewaystable}
\caption{\label{table:spe_par} Observational Parameters}
\tiny
\begin{tabular}{r|c|c|c|c|c|c|c|l} \hline \hline
Obs.\#\footnote{Full observation id is 80137-01-Obs for observations that do not start with L, 80137-02-Obs for those that start with L, denoting ``long''.} & Date & Exp.\footnote{Exposure time in ks}&\IN\ & \tin\ & PL flux\footnote{unabsorbed power-law flux in 3--25 keV band, in units of $\rm 10^{-10} \; ergs \; cm^{-2} \; s^{-1}$} & DBB flux\footnote{unabsorbed disk-blackbody flux in 3--25 keV band, in units of $\rm 10^{-10} \; ergs \; cm^{-2} \; s^{-1}$} & RMS (\%) \footnote{Rms amplitude of variability in 3-30 keV band after ridge correction.} & Notes \\
01-00 & 52908.49 &  $2.40$ & $2.73 \pm 0.07$ & $0.82 \pm 0.01$ & $5.06$ &
$24.63 $ & $<$ 3.25 & \\
02-00 & 52909.53 &  $3.00$ & $2.40 \pm 0.05$ & $0.81 \pm 0.01$ & $6.03$ &
$23.86 $ & $<$ 3.25 & \\
03-00 & 52911.52 &  $3.54$ & $2.24 \pm 0.03$ & $0.80 \pm 0.01$ & $8.34$ &
$23.00 $ & $<$ 3.23 & \\
04-00 & 52912.50 &  $3.54$ & $2.18 \pm 0.03$ & $0.800 \pm 0.01$ & $10.4$ &
$22.11 $ & $2.17 \pm 1.48$ & \\
05-00 & 52914.47 &  $2.40$ & $2.30 \pm 0.04$ & $0.780 \pm 0.01$ & $4.91$ &
$19.87 $ & $1.66 \pm 0.99$ & \\
11-00 & 52916.88 &  $3.48$ & $2.25 \pm 0.03$ & $0.76 \pm 0.01$ & $7.33$ &
$16.69 $ & $2.78 \pm 2.10$ & \\
07-00 & 52917.56 &  $3.48$ & $2.18 \pm 0.03$ & $0.74 \pm 0.01$ & $6.95$ &
$15.84 $ & $2.24 \pm 1.11$ & \\
08-00 & 52918.21 &  $2.94$ & $2.25 \pm 0.04$ & $0.75 \pm 0.01$ & $6.66$ &
$15.53 $ & $1.28 \pm 1.33$ & \\
09-00 & 52920.62 &  $2.52$ & $2.36 \pm 0.05$ & $0.74 \pm 0.01$ & $3.50$ &
$14.02 $ & $1.86 \pm 0.80$ & \\
10-00 & 52921.44 &  $2.28$ & $2.35 \pm 0.09$ & $0.73 \pm 0.01$ & $3.47$ &
$12.57 $ & $2.28 \pm 1.16$ & \\
13-00 & 52924.62 &  $2.70$ & $2.31 \pm 0.07$ & $0.70 \pm 0.01$ & $3.20$ &
$10.32 $ & $<$ 2.04 & Radio (VLA) observation with no detection \\
14-00 & 52926.18 &  $3.48$ & $2.29 \pm 0.05$ & $0.68 \pm 0.01$ & $3.70$ &
$8.65 $ & $<$ 2.53 & \\
15-00 & 52927.04 &  $3.42$ & $2.24 \pm 0.06$ & $0.68 \pm 0.01$ & $2.97$ &
$8.03 $ & $<$ 2.15& \\
16-00 & 52928.48 &  $1.86$ & $2.18 \pm 0.07$ & $0.66 \pm 0.01$ & $3.48$ &
$6.96 $ & $<$ 2.80 & Galactic ridge \wsim 10\% of the overall flux \\
17-00 & 52929.32 &  $3.54$ & $2.01 \pm 0.05$ & $0.66 \pm 0.01$ & $4.05$ &
$6.38 $ & $<$ 2.03 & \\
18-00 & 52929.91 &  $1.56$ & $2.14 \pm 0.05$ & $0.65 \pm 0.01$ & $4.58$ &
$5.87 $ & $<$ 1.37 & \\
18-01 & 52929.98 &  $2.04$ & $2.07 \pm 0.05$ & $0.65 \pm 0.01$ & $4.44$ &
$5.92 $ & $<$ 1.77 & Still in the TD state.\\ \hline
19-01 & 52930.90 &  $1.56$ & $2.20 \pm 0.04$ & $0.64 \pm 0.01$ & $8.17$ &
$4.96 $ & $7.18 \pm 0.27$  & Timing noise, transition to the IS \\
19-00 & 52931.46 &  $1.98$ & $2.27 \pm 0.03$ & $0.63 \pm 0.01$ & $8.18$ &
$4.83 $ & $7.60 \pm 0.52$ & \\
20-00 & 52932.07 &  $3.30$ & $2.22 \pm 0.02$ & $0.62 \pm 0.01$ & $11.66$ &
$3.25 $ & $11.15 \pm 0.46$ & No radio detection at MJD~52932.96. QPO at 7.8 Hz.
\\
21-00 & 52933.27 &  $3.42$ & $2.10 \pm 0.05$ & $0.62 \pm 0.01$ & $4.63$ &
$4.11 $ & $1.85 \pm 0.26$ & Timing noise reduces substantially for this obs. only! \\
22-00 & 52933.99 &  $3.48$ & $2.23 \pm 0.02$ & $0.60 \pm 0.01$ & $8.77$ &
$3.06 $ & $10.35 \pm 1.10$ & QPO at 7.6 Hz. \\
23-00 & 52935.04 &  $2.88$ & $2.25 \pm 0.03$ & $0.58 \pm 0.01$ & $6.07$ &
$2.89 $ & $9.77 \pm 1.36$ & QPO at 7.1 Hz. \\
24-00 & 52936.02 &  $2.28$ & $2.20 \pm 0.02$ & $0.58 \pm 0.01$ & $7.88$ &
$2.12 $ & $11.44 \pm 0.56$ & QPO at 7.7 Hz. \\
25-00 & 52937.51 &  $3.48$ & $2.17 \pm 0.02$ & $0.55 \pm 0.02$ & $9.47$ &
$1.26 $ & $15.71 \pm 0.91$ & QPO at 6.9 Hz. \\
26-00 & 52938.00 &  $2.10$ & $2.06 \pm 0.02$ & $0.52 \pm 0.03$ &
$9.61$ & $0.80 $ & $21.22 \pm 1.13$ & LHS according to
\citep{McClintock03}. QPO at
5.9 Hz. \\
27-00 & 52939.12 &  $2.04$ & $2.07 \pm 0.02$ & $0.50 \pm 0.03$ & $8.58$ &
$0.67 $ & $19.47 \pm 0.78$ & Radio (VLA) detection! QPO at 5.8 Hz. \\
28-00 & 52942.12 &  $1.98$ & $1.86 \pm 0.02$ & $0.50 \pm 0.08$ & $6.31$ &
$0.20 $ & $26.21 \pm 1.08$ & QPO at 2.9 Hz. \\
29-00 & 52942.68 &  $1.62$ & $1.87 \pm 0.03$ & $0.59 \pm 0.11$ & $5.66$ &
$0.19 $ & $19.82 \pm 2.34$ & QPO at 2.4 Hz. \\
L01-00 & 52944.10 &  $10.08$ & $1.78 \pm 0.01$ & $0.42 \pm 0.05$ & $4.91$ &
$0.13 $ & $27.51 \pm 2.63$ & QPO at 1.8 Hz. \\
30-00 & 52945.17 &  $1.80$ & $1.78 \pm 0.02$ & $0.45 \pm 0.06$ & $4.02$ &
$0.14 $ & $26.46 \pm 3.69$ & \\
L01-01 & 52946.07 &  $10.62$ & $1.69 \pm 0.02$ & $0.47 \pm 0.07$ & $3.59$ &
$0.09 $ & $28.96 \pm 2.54$ & \\
31-00 & 52947.02 &  $1.08$ & $1.79 \pm 0.03$ & $0.39 \pm 0.10$ & $2.80$ &
$0.06 $ & $29.93 \pm 5.11$ & \\
32-00 & 52948.19 &  $5.22$ & $1.78 \pm 0.02$ & $0.29 \pm 0.02$ & $2.29$ &
$0.03 $ & $30.11 \pm 2.23$ & \\
33-00 & 52949.53 &  $3.54$ & $1.74 \pm 0.02$ & $0.37 \pm 0.07$ & $1.99$ &
$0.05 $ & $36.40 \pm 2.41$ & Last radio (VLA) detection. \\
L01-02 & 52949.67 &  $10.38$ & $1.76 \pm 0.01$ & $0.28 \pm 0.01$ & $1.79$ &
$0.03 $ & $39.29 \pm 5.45$ &\\
34-00 & 52950.44 &  $2.04$ & $1.78 \pm 0.03$ & $0.36 \pm 0.06$ & $1.47$ &
$0.03 $ & $32.27 \pm 12.20$ & Only PCA is used from this observation on. \\
35-00\footnote{34-01,35-00 and 35-01 were merged} & 52951.39 &  $3.72$ & $1.74 \pm 0.03$ & $0.37 \pm 0.06$ & $1.02$ & $0.04 $ & - & Galactic ridge is as strong
as the source. \\
36-00 & 52952.26 &  $1.56$ & $1.86 \pm 0.04$ & $0.29 \pm 0.03$ & $0.84$ &
$0.01 $ & - & \\
L02-00 & 52953.18 &  $14.28$ & $1.83 \pm 0.02$ & $0.28 \pm 0.02$ & $0.58$ &
$0.02 $ & - & \\
L02-01 & 52953.69 &  $3.96$ & $1.83 \pm 0.05$ & $0.30 \pm 0.04$ & $0.49$ &
$0.02 $ & - & \\
L02-02 & 52953.96 &  $8.76$ & $1.72 \pm 0.05$ & $0.40 \pm 0.03$ & $0.41$ &
$0.04 $ & - & \\
37-00 & 52955.46 &  $1.74$ & $1.98 \pm 0.22$ & $0.38 \pm 0.08$ & $0.18$ &
$0.03 $ & - & \\
38-00 & 52956.26 &  $3.06$ & $1.71 \pm 0.20$ & $0.35 \pm 0.06$ & $0.13$ &
$0.02 $ & - & Radio (ATCA) observation with no detection. \\
39-00 & 52957.00 &  $1.44$ & $2.43 \pm 0.29$ & $0.36 \pm 0.09$ & $0.09$ &
$0.02 $ & - & \\
40-00 & 52958.23 &  $3.66$ & $2.32 \pm 0.25$ & $0.33 \pm 0.06$ & $0.08$ &
$0.02 $ & - & \\
L03-0x\footnote{L03-00, L03-02, L03-03 were merged.} & 52958.82 &  $12.12$ & $2.51 \pm 0.66$ & $0.35 \pm 0.05$ & $0.02$ &
$0.02 $ & - & \\
L03-01 & 52959.21 &  $14.64$ & $1.79 \pm 0.41$ & $0.36 \pm 0.03$ & $0.03$ &
$0.02 $ & - & \\
\end{tabular}
\end{sidewaystable}

\clearpage

\end{document}